\documentclass[a4paper]{jpconf}
\usepackage{graphicx}

\usepackage{euscript}

\usepackage{latexsym}

\def\mn#1#2{\ensuremath{|\Delta M_{#1#2}^2}|}

\renewcommand{\thefootnote}{\alph{footnote}}
\renewcommand{\(}{\begin{equation}}
\renewcommand{\)}{\end{equation}}
\newcommand{\ba}{\begin{eqnarray}}
\newcommand{\ea}{\end{eqnarray}}

\newcommand{\bp}{\mathop{\vtop{\ialign{##\crcr
   $\hfil\displaystyle{}\hfil$\crcr\noalign{\kern-13pt\nointerlineskip}
   \BIG{(}\hskip0pt\crcr\noalign{\kern3pt}}}}}
\newcommand{\cbp}{\mathop{\vtop{\ialign{##\crcr
   $\hfil\displaystyle{}\hfil$\crcr\noalign{\kern-13pt\nointerlineskip}
   \BIG{)}\hskip0pt\crcr\noalign{\kern3pt}}}}}
\newcommand{\pa}{\mathop{\vtop{\ialign{##\crcr
    
$\hfil\displaystyle{\oplus}\hfil$\crcr\noalign{\kern+1pt\nointerlineskip 
}
   \hspace{.08in}$^{\alpha=0}$\hskip6pt\crcr\noalign{\kern3pt}}}}}

\def\mn#1#2{\ensuremath{|\Delta M_{#1#2}^2}|}

\def\dc{\Delta\chi^2}
\def\odc{\overline{\dc}}

\newcommand{\beq}{\begin{equation}}
\newcommand{\eeq}{\end{equation}}
\newcommand{\bea}{\begin{eqnarray}}
\newcommand{\eea}{\end{eqnarray}}
\newcommand{\beas}{\begin{eqnarray*}}
\newcommand{\eeas}{\end{eqnarray*}}

\newcommand{\bquo}{\begin{quote}}
\newcommand{\enqu}{\end{quote}}

\begin{document}
\title{Confidence in the neutrino mass hierarchy}
\author{Jarah Evslin}
\address{Theoretical Physics Division, IHEP, 
CAS, YuQuanLu 19B, Beijing 100049, China }
\ead{jarah@ihep.ac.cn}
\begin{abstract}
 The number of $\sigma$ of confidence in a determination of the neutrino mass hierarchy may be obtained from the statistic $\overline{\Delta\chi^2}$.  However,  as the hierarchy is a discrete variable, this number is not simply $\sqrt{\overline{\Delta\chi^2}}$.  We review a simple Bayesian formula for the sensitivity to the hierarchy determination that can be obtained from the median experiment as a function of $\overline{\Delta\chi^2}$.  We compare this analytical formula to 6 years of simulated data from JUNO together with a 4\% (1\%) determination of the effective atmospheric mass splitting from the disappearance channel at MINOS (NO$\nu$A).  We find a $\overline{\Delta\chi^2}$ of 11 (20) which we identify with $2.6\sigma$ ($3.9\sigma$).   However when the unknown nonlinear energy response of the detector is included in our analysis this significance degrades considerably.  This degradation can be eliminated by dividing the single detector into a near and far detector of the same total target mass.  A further advantage of a second detector is that, even while the reactor neutrino experiment runs, the decay at rest of a {\it{single}}, high intensity, continuously running $\pi^+$ source close to one of the detectors, such as that described by the DAE$\delta$ALUS project, may determine the leptonic CP-violating phase $\delta$.

\end{abstract}

\section{Measures of sensitivity to the hierarchy}
\subsection{The $\odc$ statistic}
The sensitivity to the neutrino mass hierarchy can be obtained from the statistic
\beq
\Delta\chi^2=\chi^2_I-\chi^2_N \label{deltanoi}
\eeq
where $\chi^2_N$ ($\chi^2_I$) is the $\chi^2$ statistic for a fit of the data $y_i$ to the expected data $y_i^N(x)$ ($y_i^I(x)$) given the normal (inverted) hierarchy separately minimized over the nuisance parameters $x$
\beq
\chi_{(I,N)}^2=\inf_x \sum_i\frac{(y_i-y_i^{(I,N)}(x))^2}{\sigma_i(x)^2}.
\eeq

Note that Eq. (\ref{deltanoi}) is not the familiar formula for $\Delta\chi^2$ from the Particle Data Book, which is equal to the $\chi^2$ statistic for a hypothesis minus that for the best fit.  The familiar formula is always nonnegative because a best fit by definition minimizes $\chi^2$, and in fact Wilks' theorem implies that it follows a one degree of freedom $\chi^2$ distribution, from which one can prove that the hypothesis is excluded with a significance of $\sqrt{\Delta\chi^2}\hspace{0.1cm}\sigma$.

On the other hand the statistic $\Delta\chi^2$ defined in Eq. (\ref{deltanoi}) is the difference between the $\chi^2$ value for two hypotheses and so can be negative, thus it cannot follow a $\chi^2$ distribution and $\sqrt{\Delta\chi^2}$ is not even necessarily real.  In Ref \cite{xinoct} the authors demonstrated that in the case with no nuisance parameters $\Delta\chi^2$ follows a Gaussian distribution with mean $\odc$ and standard deviation
\beq
\sigma=2\sqrt{\odc}.
\eeq
In Ref. \cite{noistat} this result was extended to experiments with nuisance parameters, where it was found that it applies when a $1\sigma$ variation in the nuisance parameters yields a variation in $\odc$ which is smaller than $\sigma$.  The authors checked that this condition is satisfied for RENO 50 and JUNO.  In fact, it appears to be the case for accelerator, atmospheric and reactor experiments designed to determine the neutrino mass hierarchy except for T2K and NO$\nu$A, where $\delta$ provides a nuisance parameter whose 1$\sigma$ variation can, depending upon its value, reduce $\odc$ to zero.  However, once $\delta$ is determined, this result may be expected to apply to T2K and NO$\nu$A as well. 

\subsection{Sensitivity to the hierarchy}
$\odc$ can be determined via simulations before an experiment begins.  Given a value of $\odc$, what is the expected sensitivity of the experiment to the hierarchy?  There are several questions here, all of which have been answered in Ref. \cite{noistat}.  In what follows we will repeat these answers in a Bayesian approach in which the prior probability of each hierarchy is $50$ percent.
\vspace{.1cm}

\noindent
{\it{First, what is the probability that the hierarchy which yields the lowest $\chi^2$ is indeed the true hierarchy?}}
The probability of correctly determining the hierarchy is
\beq
p_c(\odc)=\frac{1}{2}\left(1+{\rm{erf}}\left(\sqrt{\frac{\overline{\Delta\chi^2}}{8}}\right)\right). \label{suc}
\eeq
This is the quantity quoted in a number of studies such as Refs.~\cite{caojun2,xinag,noisim,whitepaper,bari}.  It corresponds to a sensitivity which is roughly half of the number of $\sigma$ suggested by Wilks' theorem.

\vspace{.1cm}

\begin{figure} 
\begin{center}
\includegraphics[width=2.8in,height=1.3in]{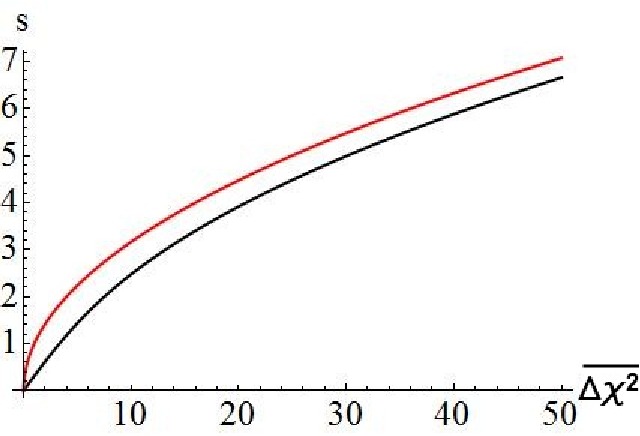}
\includegraphics[width=2.8in,height=1.3in]{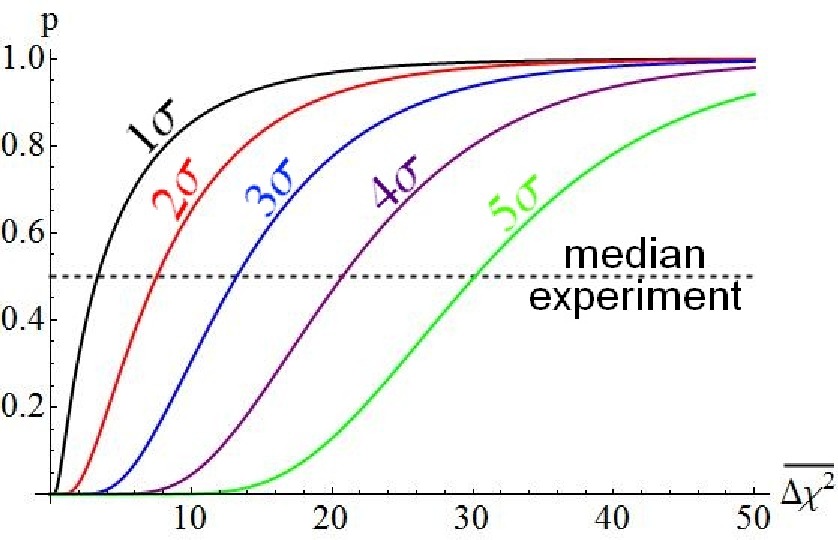}
\caption{{\bf{Left:}} The black curve is the number $s$ of $\sigma$ of sensitivity of the determination of the mass hierarchy by a median experiment. The red curve is the square root of $\overline{\Delta{\chi^2}}$. {\bf{Right:}} The black, red, blue, purple and green curves are the probability of a hierarchy determination with $1\sigma$, $2\sigma$, $3\sigma$, $4\sigma$ and  $5\sigma$ of sensitivity.  The dashed line represents a median experiment.}
\label{sfig}
\end{center}
\end{figure}

\noindent
{\it{Second, with what is the sensitivity to the hierarchy of a typical experiment?}}
A ``typical experiment" is one in which $|\Delta\chi^2|=|\overline{\Delta\chi^2}|$.  This corresponds to the median value of the probability of success.   The probability that a fit to the correct hierarchy yields a lower value of $\chi^2$ than one to the wrong hierarchy is simply the probability that $\Delta\chi^2$ has the correct sign
\beq
p_v=\frac{1}{1+e^{-\overline{\Delta\chi^2}/2}}\label{princ}
\eeq
corresponding to
\beq
s(\overline{\Delta\chi^2})=\sqrt{2}\ {\rm{erf}}^{-1}\left(\frac{1-e^{-\overline{\Delta\chi^2}/2}}{1+e^{-\overline{\Delta\chi^2}/2}}\right)\label{s}
\eeq
$\sigma$ of sensitivity, as is  plotted in the left panel of Fig.~\ref{sfig}.  For example, if $\overline{\Delta\chi^2}=9$ then the probability that a median experiment correctly determines the hierarchy will be 98.9\%.  While this is better than the mean probability of success 93.3\% of Eq.~( \ref{suc}), it still falls noticeably short of the 99.7\% of sensitivity which one might expect from Wilks' theorem.

\vspace{.1cm}

\noindent
{\it{Third, what is the probability $p(s)$ that the sensitivity to the hierarchy will be at least $s\sigma$?}}
The answer, plotted in the right panel of Fig.~\ref{sfig}, is 
\beq
p(s)=\frac{1}{2}\left(1+{\mathrm{erf}}\left(\frac{\overline{\Delta\chi^2}-{\mathrm{arctanh}}\left({\mathrm{erf}}\left(\frac{s}{\sqrt{2}}\right)\right)}{\sqrt{8\overline{\Delta\chi^2}}}\right) \right) .
\eeq

\section{Synergy between reactor and accelerator disappearance experiments}

\begin{figure} 
\begin{center}
\includegraphics[width=2.8in,height=1.2in]{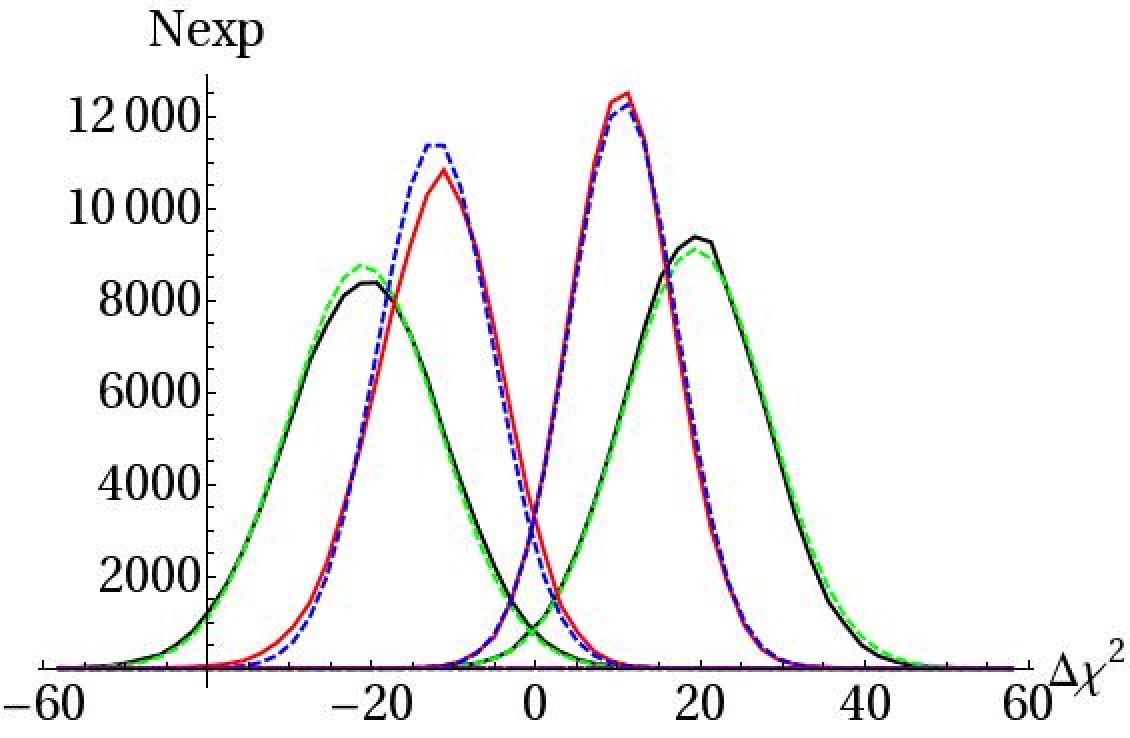}
\includegraphics[width=2.8in,height=1.2in]{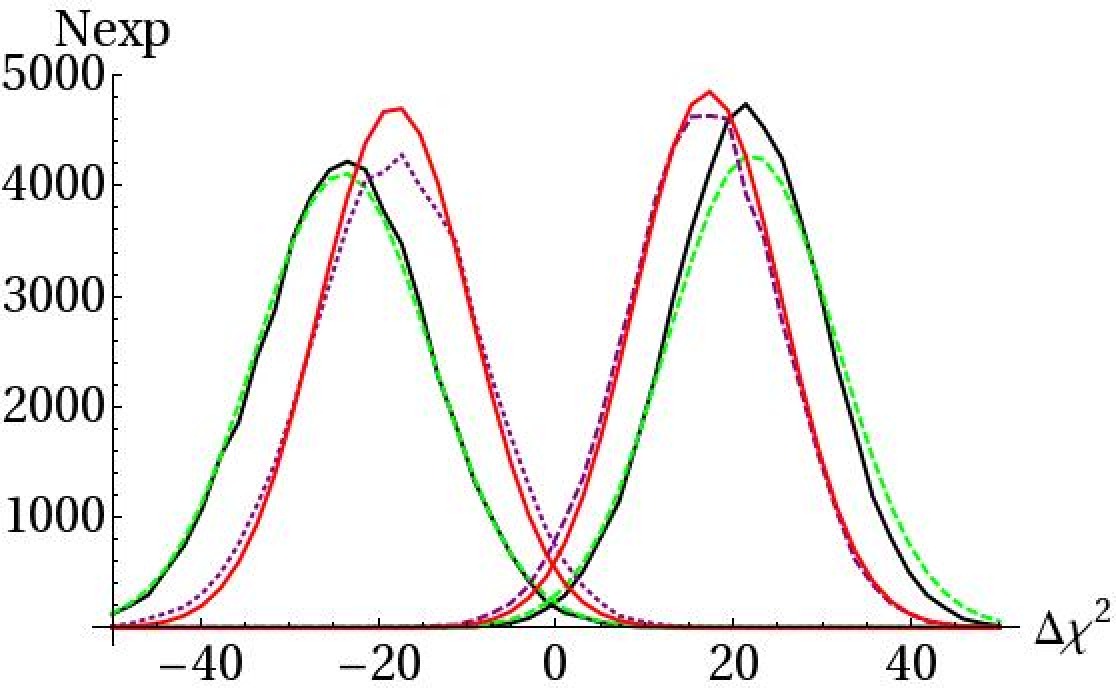}
\caption{{\bf{Left:}} The distribution of $\Delta\chi^2$ combining six years of running at JUNO with a 4\% determination of the atmospheric mass splitting (red curve) and an optimistic 1\% determination  (black curve).  The dashed curves are the Gaussian distributions centered at $\overline{\Delta\chi^2}$ with width $2\sqrt{\overline{\Delta\chi^2}}$. {\bf{Right:}} Using a 1\% determination of the atmospheric mass splitting,  the red and black curves use $\delta=0$ and $\pi$ respectively,  fit  assuming $\delta=\pi/2$.}
\label{NOvAfig}
\end{center}
\end{figure}

As an example, in the left panel of Fig.~\ref{NOvAfig} we present the distribution of the $\Delta\chi^2$ statistic in 100,000 simulations which combine the $\overline{\nu}_e$ spectrum measured in 6 years at JUNO with MINOS' 4\% determination of the atmospheric mass difference and with an optimistic 1\% forecast determination at an upgraded NO$\nu$A .    $\mn32$ is chosen to minimize $\chi^2_I$ and $\chi^2_N$.   The baselines and reactor fluxes are identical to Ref.~\cite{yifangmar}.  The leptonic CP-violating angle $\delta$ is set to $\pi/2$.   

$\overline{\Delta\chi^2}= 11\ (20)$ for JUNO with MINOS (NO$\nu$A) yielding $2.6\sigma$ ($3.9\sigma$) of sensitivity at the median experiment,  determining the hierarchy with probability 94.6\% (98.5\%) in agreement with Eq.~(\ref{suc}).    In the right panel of Fig.~\ref{NOvAfig} we present the distribution of $\Delta\chi^2$ in simulations in which $\delta=0$ and $\pi$, although we always fit to a $\delta=\pi/2$ theoretical model.   At $\delta=0$ ($\pi$) we find $\overline{\Delta\chi^2}=17$ (22) yielding $3.5\sigma$ (4.2$\sigma$) of sensitivity, confirming the expectations of Ref.~\cite{minakata}.  

\section{The two detector proposal}
The determination of the mass hierarchy at a reactor experiment requires a determination of the absolute energy scale at an unprecedented precision.  A scintillator detector determines the energy of a neutrino by counting photoelectrons, but the relationship between this number and the energy is nonlinear and will not be known precisely even after the calibration campaign.

\begin{figure} 
\begin{center}
\includegraphics[width=2.6in,height=1.3in]{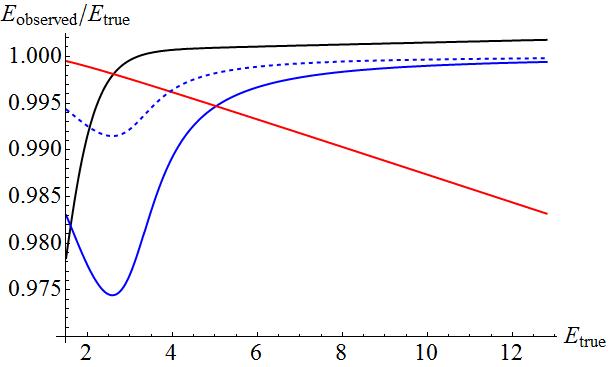}
\includegraphics[width=2.6in,height=1.3in]{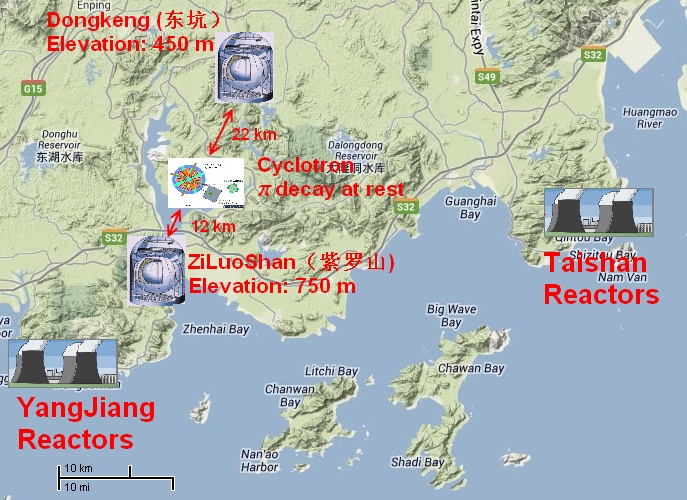}
\caption{{\bf{Left:}} Models of nonlinearity considered in Ref.~\cite{noinonlin}.  We consider the blue dashed line. {\bf{Right:}} Proposed near detector and cyclotron $\pi^+$ decay at rest source locations for JUNO}
\label{twofig}
\end{center}
\end{figure}

In Ref.~\cite{noinonlin} various models of the remaining unknown nonlinear energy response, shown in the left panel of Fig.~\ref{twofig}, were included in simulations of JUNO and RENO 50.  In the case of the dashed blue line of this figure, corresponding to an optimistic estimate of the unknown energy response, the effect of the energy response on $\odc$ is summarized in Table~\ref{ristab} for various detector locations.  Configurations with both a single detector and two half-mass detectors are considered.  As can be seen from this table, the use of two detectors at sufficiently different baselines can eliminate the reduction in $\odc$ caused by the unknown energy response.
\begin{table}
\caption{\label{ristab}$\odc$ obtained at various JUNO (RENO 50) sites after 6 years of running considering a (im)perfect energy response with single 20 kton (18 kton) and pairs of 10 kton (9 kton) detectors.}
\begin{center}
\begin{tabular}{cllll}
\br
Site&NH:No Nonlin&IH: No Nonlin&NH: Nonlin&IH: Nonlin\\
\mr
DongKeng&14.1&-17.0&8.2&-21.5\\
DongKeng+LuGuJing&13.2&-16.2&7.8&-21.4\\
DongKeng+ZiLuoShan&13.5&-16.1&13.9&-15.3\\
GuemSeong&6.2&-7.7&3.3&-10.0\\
GuemSeong+Jangamsan&5.6&-6.6&5.3&-7.0\\
Munmyeong&11.8&-13.6&6.4&-17.5\\
Munmyeong+Buncheon-ri&11.5&-13.6&9.4&-16.4\\
\br
\end{tabular}
\end{center}
\end{table}

Another advantage of including a second detector is that one may, similarly to the DAE$\delta$ALUS project \cite{daed}, determine $\delta$ in a $\pi^+$ decay at rest experiment.  However, with multiple detectors only a single cyclotron pair is required and it may run 100\% of the time.

ZiLuoShan is high enough to afford significant protection from cosmic muons.  Yet at 700 meters of depth, JUNO's preferred DongKeng site can expect 5 muons/second, of which more than 10\% may be showering (energy deposition of 3 GeV beyond that from ionization) and a similar fraction of muon bundles.  Thus KamLAND's 2 second full detector veto cannot be applied.  Without cuts one expects 200,000 ${}^9$Li decays in 6 years, outnumbering the $\nu$ signal.  It is not clear whether smarter cuts may suffice or whether JUNO must be dug deeper.

\ack
JE is supported by the CAS
Fellowship for Young International Scientists grant 
2010Y2JA01. 

\section*{References}

\end{document}